\documentclass[]{spie}  

\usepackage{amsmath,amsfonts,amssymb}
\usepackage{graphicx}
\usepackage[colorlinks=true, allcolors=blue]{hyperref}

\title{Pollux: decisions affecting the optical architecture of a high-resolution spectrograph and polarimeter for the Habitable Worlds Observatory  }

\author[a,d] {Eduard Muslimov} 
\author[b] {Luca Fossati} 
\author[c] {Coralie Neiner} 
\author[d] {Jean-Claude Bouret}
\author[d] {David Le Mignant}
\author[d] {Kjetil Dohlen}
\author[c] {Jean-Michel Reess}
\author[c] {Adrien Girardot}

\affil[a]{Department of Physics, University of Oxford, Keble Rd, OX14 3RH Oxford, UK}
\affil[b]{Space Research Institute, Austrian Academy of Sciences, Schmiedlstrasse 6, 8042 Graz, Austria}
\affil[c]{LIRA, Observatoire de Paris, Universite PSL, CNRS, Sorbonne Universite, Universite Paris Cite, CY Cergy Universite, 92190 Meudon, France}
\affil[d]{Aix Marseille Univ, CNRS, CNES, LAM, Marseille, France}

\authorinfo{Further author information: (Send correspondence to E.M.)\\E.M.: E-mail: eduard.muslimov@physics.ox.ac.uk}

\begin{document} 
\maketitle

\begin{abstract}
POLLUX is a candidate European instrumental contribution to the Habitable Worlds Observatory. It is a high-resolution spectrograph with polarimetric capabilities, covering from the far ultraviolet (FUV; $100 nm$) to the near infrared (NIR; $1.75 mu m$). Such a broad spectral coverage is achieved by splitting the instrument into five channels, each comprising an echelle spectrograph: FUV, medium-UV (MUV), near-UV (NUV), optical (OPT), and NIR. A set of custom-made dichroics enables simultaneity across the MUV, NUV, OPT, and NIR channels. We present the latest developments in the optical design of the three UV channels. Specifically, we estimate the impact of telescope residual jitter on resolving power and sampling and discuss possible options to enable pure spectroscopy in the FUV channel without implementing a fully retractable polarimeter and to compensate the defocus when inserting MUV and NUV polarimeters. Finally, we estimate the impact of detector size limitation and potential advantages of shrinking or extending the wavelength coverage in the NUV channel.
\end{abstract}


\section{INTRODUCTION}
\label{sec:intro} 
The Habitable Worlds Observatory (HWO) should become the first telescope designed to search for extraterrestrial life, but it also should serve for multiple science cases in astrophysics. After the recommendations of the Astro2020 Decadal Survey for large missions the HWO project entered the Technology Maturation Phase in 2024.  The studies for the future telescope payload are being carried out in parallel and should find the best synergies between the unique capabilities of the telescope and performance of the separate instruments onboard \cite{Feinberg26}. 

Such unique properties intrinsic to the HWO project such as its unprecedented collecting power, extremely high pointing stability and transmission optimization over a wide spectral range became driving factors for the Pollux instrument. Pollux is a candidate European instrumental contribution to HWO. It is a high-resolution spectrograph, with spectropolarimetric capabilities, covering from the FUV to the NIR\cite{Neiner26}. It is proposed to complement other instruments currently considered for HWO, i.e. a coronagraph, a high-resolution imager, a UV multi-object spectrograph (MOS), and a UV integral field unit (IFU).
The driving Pollux science cases have already been identified \cite{Neiner26}, and thus it is possible to define the top-level performance requirements (see Sec.~\ref{sec:REQ}). The baseline instrument architecture has also been set \cite{LeMignant26} and fundamental properties such as material transmission and reflection curves, or grating diffraction law, give the basic instrumental capabilities. A brief overview is given in Sec.~\ref{sec:BASIC}. Instead, the interfaces with telescope and some main technological risks remain open. Therefore, in this work we discuss a few decisions in the instrument optical design that can make a notable impact on its operation modes and performance. In particular, the following questions are of a primary importance:
\begin{itemize}
    \item The telescope pointing stability should be high, but the exact value is not known yet. The residual jitter will define the point spread function (PSF) shape at the instrument entrance focal plane. It is thus necessary to study the impact of pointing stability on instrumental performance (see Sec.~\ref{sec:JITTER}).
    \item The instrument design and development would benefit from the use of single-chip detectors. Among other things, this would avoid blind zones, increase instrumental stability and simplify interfaces. However, this will limit the linear size of the spectrogram and affect the choice of camera and dispersers. Verifying the impact of this limitation is another goal of the present paper (see Sec.~\ref{sec:JITTER}).
    \item Maximize the effective area is a major design driver, particularly in the UV. One solution for spectroscopy-only observations involves retracting or by-passing the polarimeters. However, the implementation of retractable polarimeters introduces the need to refocus the beam following the insertion of the polarimeter. We discuss way of achieving this without the implementation of complex refocusing mechanisms (Sec.~\ref{sec:POLAR}). 
\end{itemize}

\section{TOP-LEVEL REQUIREMENTS}
\label{sec:REQ}  
 \begin{table}[ht]
\caption{\label{tab:top_specs} Top-level specifications of the Pollux channels.}
\begin{center}
\begin{tabular}{l c c c c c}
\hline
\rule[-1ex]{0pt}{3.5ex} Metric & FUV & MUV & NUV & OPT & NIR \\
\hline
\rule[-1ex]{0pt}{3.5ex} Wavelength range (nm) & 100--123 & 120--236 & 236--438 & 438--875 & 875--1750 \\
\hline
\rule[-1ex]{0pt}{3.5ex} Spectral resolving power & 100k & 100k & 100k & 65k & 65k \\
\hline
\rule[-1ex]{0pt}{3.5ex} Point source spectropolarimetry & Yes & Yes & Yes & Yes & Yes \\
\hline
\rule[-1ex]{0pt}{3.5ex} Point source spectroscopy & No & Yes & Yes & Yes & Yes \\
\hline
\rule[-1ex]{0pt}{3.5ex} Slit spectroscopy & No & Yes & Yes & No & No \\
\hline
\rule[-1ex]{0pt}{3.5ex} Simultaneity & No & Yes & Yes & Yes & Yes \\
\hline
\end{tabular}
\end{center}
\end{table}
The key target specifications of the instrument are summarized in Table~\ref{tab:top_specs}. The primary aim is to maximize spectral coverage and simultaneity. The lower spectral limit is defined mainly by the telescope coating properties. With $LiF$-based coatings we can expect a sufficient transmission above $100 ~nm$\cite{Feinberg26}, although depending on some other designs solutions there is a possibility to use the secondary maximum of the reflectivity curve and extend the working range down to $97~nm$. The properties and characteristics of the FUV channel are constrained by the available choice of materials -- there are practically no known birefringent transmissive materials at these wavelengths, polarization modulation is challenging and the choice of both reflective coatings and detectors design are limited. This is why the FUV channel is separated from the others. The upper wavelength limit is set in such a way to guarantee that a broaden, non-shifted $Ly\alpha$ line is always fully covered. The original primary mode for the FUV channel is point-source spectropolarimetry, which covers most of the science cases\cite{Neiner26}, but this limits significantly the throughput, which is why we explore here ways to by-pass the polarimeter to boost the effective area for spectroscopy-only observations.

The MUV and NUV channels have working spectral bands built up from the FUV upper wavelength boundary and considering the lower working limit of 120 nm for birefringent materials. Each channel must cover in wavelength less than octave to avoid spectra overlapping after diffraction. In both channels the polarimeter can be based on birefringent crystals, which enables the possibility to retract them for spectroscopy-only observations to boost throughput. In spectroscopy-only mode one gains space on the detector, which in polarimetric mode is used to record the parallel an perpendicular beams, enabling the implementation of a slit-mode (slit length of $\sim$3") to observe extended sources. The spectral resolving power in the UV channels is still under discussion, but the majority of the driving science cases requires a value $\approx100\,000$.

To optimize the space available in the instrument bay the OPT and NIR channels are currently connected to the rest of the instrument through dedicated optical fibers. Each channel has a dedicated retractable polarimeter, making spectropolarimetry an optional mode. The driving science cases require a somewhat lower resolving power of $\approx$65\,000 compared to the UV channels. This makes the OPT and NIR channels more manageable and efficient. However, the fiber link connection leads to limitations (e.g., no slit mode) and possible problems (e.g., modal noise, coarse flux stability), and thus this choice will be revisited once final volume information become available and a full mechanical and thermal analysis can be carried out.


\section{INTERFACE WITH TELESCOPE AND BASIC DESIGN}
\label{sec:BASIC}  

\begin{figure} [ht]
   \begin{center}
   \begin{tabular}{c} 
   \includegraphics[width=0.95\textwidth]{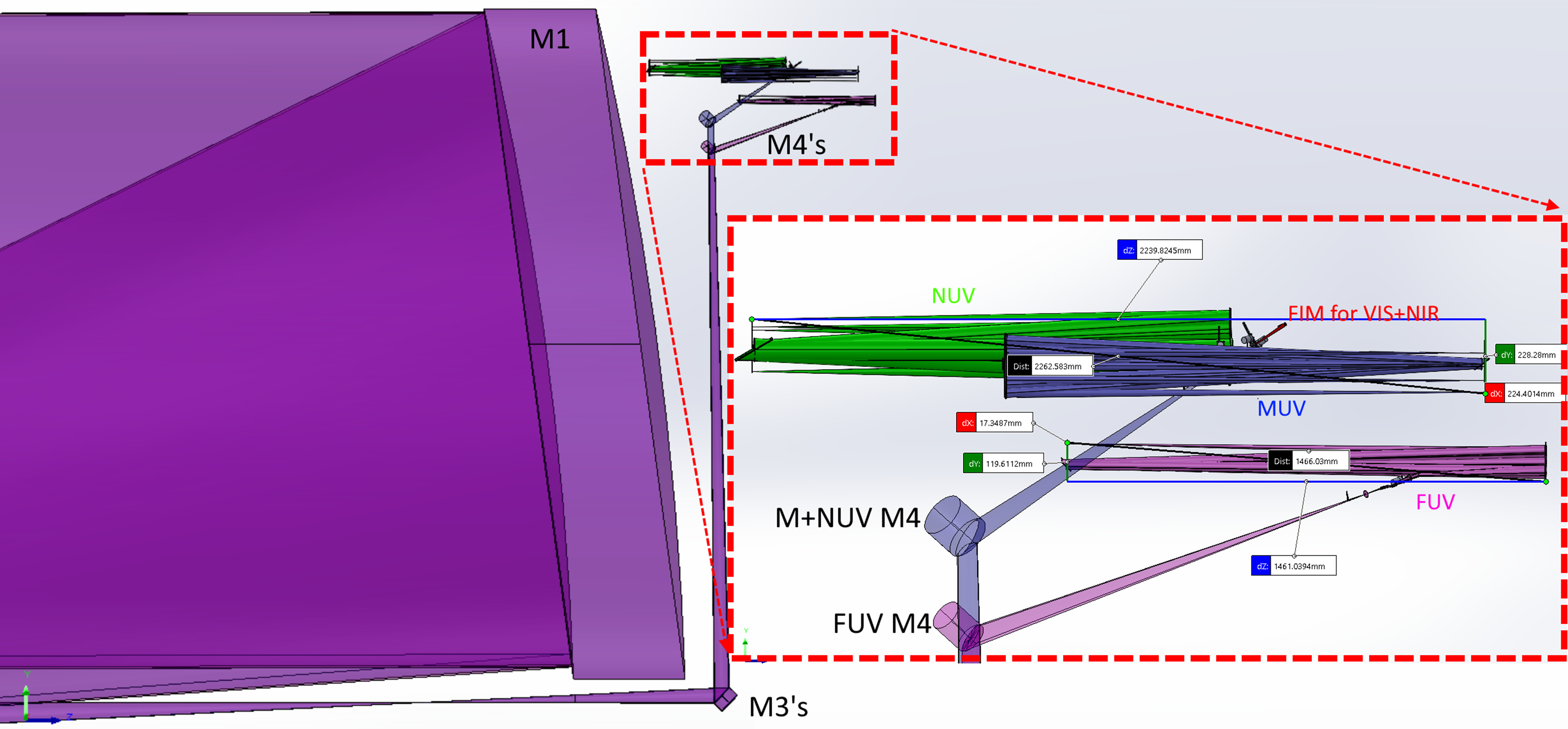}
   \end{tabular}
   \end{center}
   \caption{\label{fig:telescope interface} Possible implementation of the Pollux optical interface with EAC5 based on two customized \textit{M3-M4} relays. The OPT and NIR channels are not shown.}
   \end{figure}

At the current development stage, two second-generation Early Architecture Concepts (EACs) for the HWO telescope are under consideration \cite{alice2026} and three EACs were analyzed at the previous stage. The latest concepts comprise a telescope optical system consisting of four mirrors, where the first two mirrors \textit{M1} and \textit{M2} are common to all payload instruments, while the \textit{M3-M4} relays will be individual to each of them. Coupling of Pollux with a version of the telescope approximately corresponding to EAC5 is shown in Fig.~\ref{fig:telescope interface}.           

No dichroic splitting FUV photons and meeting the reflection and transmission requirements can be currently made, therefore the FUV channel has to work separately from the others. One possible option for sending light into the FUV channel is using a flip mirror, but this implies the introduction of another bounce and precise mechanism. The better alternative is to use a separate relay with optimized \textit{M3-M4} mirrors. In this case, switching to the FUV channel would be achieved by a simple re-pointing of the telescope. Fortunately, there is enough space for two mirror relays behind the primary mirror. As a further important benefit of this solution is that we can change the optical power of each mirror independently to shift the FUV and MUV--NIR focal points varying the corresponding $F/\#'s$, if necessary. 

\begin{figure} [ht]
   \begin{center}
   \begin{tabular}{c} 
   \includegraphics[width=0.9\textwidth]{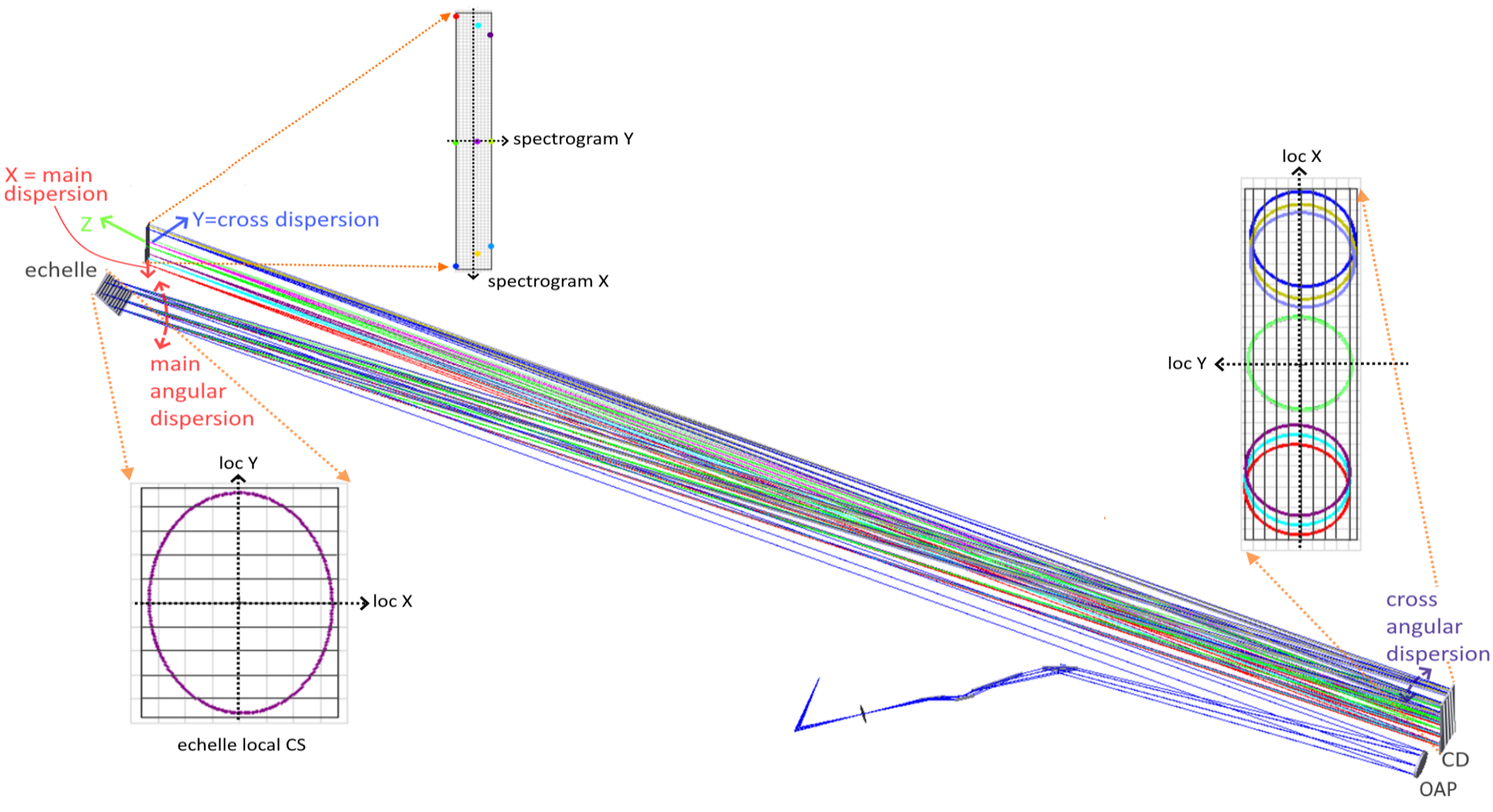}
   \end{tabular}
   \end{center}
   \caption{\label{fig:fuv layout}  Optical design of the FUV echelle spectropolarimeter and definition of the common axes notation.}
\end{figure}

With the given target resolving power, the working spectral range and the space environment there is no alternative to an echelle spectrograph for each channel. In general, the design remains similar to that of previous iterations of the project \cite{Muslimov24}. Fig.~\ref{fig:fuv layout} shows the optical design of the FUV channel as an example. 

\begin{figure} [ht]
   \begin{center}
   \begin{tabular}{c} 
   \includegraphics[width=0.45\textwidth]{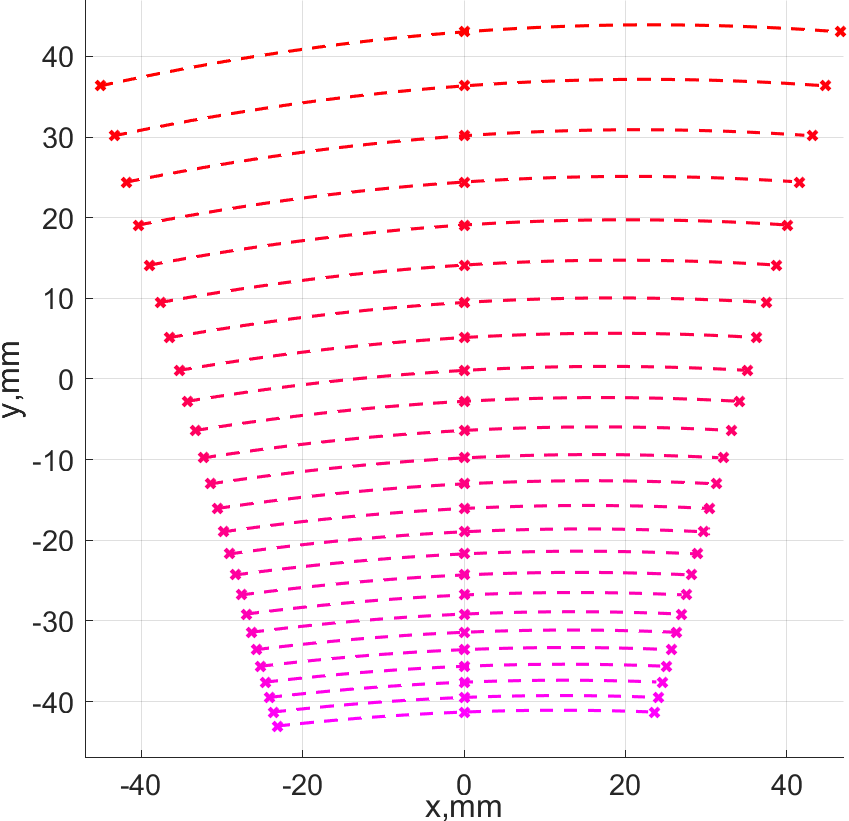}
   \end{tabular}
   \end{center}
   \caption{\label{fig:echellogram}  Example of the 2D MUV spectrogram in pure spectroscopy mode.}
   
   \end{figure}

The telescope beam is focused on a pinhole. In the case of the FUV channel, the beam passes through a dedicated polarimeter module, comprising a K-mirror rotating modulator and Brewster angle analyzer, to be then re-collimated by an off-axis paraboloid (OAP) mirror. The collimated beam is dispersed by an echelle grating and then the working orders are separated by the cross-disperser (CD) grating. To minimize reflections, the cross-disperser is concave to also perform the camera function. The 2D spectrogram is focused on a detector plane. Other channels follow the same basic design, although the polarimetric units are different and fully retractable (see Sec.~\ref{sec:POLAR} and works by Neiner and Girardot\cite{Neiner25, Girardot24}). 
Furthermore, the number of echelle orders, their non-linear separation and curvature increase at longer wavelengths, with the orders separation being limited by the requirement of separating the ordinary and extraordinary beams in spectropolarimetric mode The clocking of the spectrogram is defined by the \textit{Z} rotation of the echelle grating that de-tunes it from the exact Littrow condition and allows to avoid the collimator and camera collision. Fig.~\ref{fig:echellogram} shows these effects for the MUV channel when projecting the entire waveband onto a single detector. 

\section{SAMPLING TRADE-OFF AND JITTER EFFECT}
\label{sec:JITTER}  

\begin{figure} [ht]
   \begin{center}
   \begin{tabular}{c} 
   \includegraphics[width=0.9\textwidth]{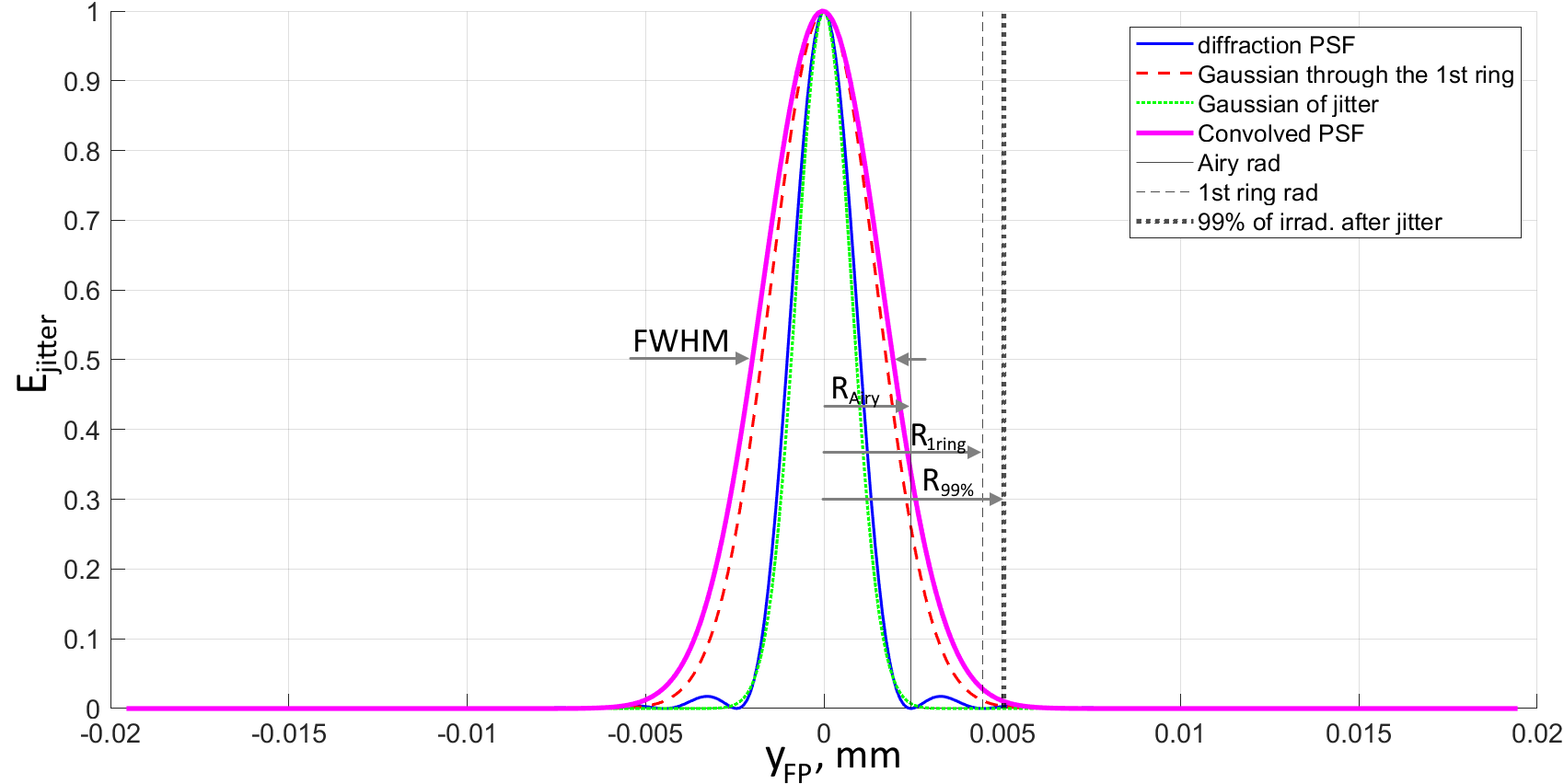}
   \end{tabular}
   \end{center}
   \caption{\label{fig:jitter_conv}  Simplified representation of the telescope PSF cross-section including jitter effects. } 
\end{figure}

The HWO telescope is expected to be diffraction-limited in the visible and its pointing stability will be extremely high. The exact values are not known yet, but they will affect the input PSF of Pollux, and thus the instrument design and performance. In order to quantify these effects we make the following assumptions (see Fig.~\ref{fig:jitter_conv} for an illustration).

We assume that the residual aberrations will be relatively small, so $>99\%$ of energy will be concentrated within the first diffraction ring. Therefore, the aberrated PSF diameter is approximated with a gaussian curve, defined by 
  \begin{equation}
\label{eq:1ring}
R_{1 ring}  = 3\sigma=\frac{4.46 \lambda f'_{tel}}{D_{tel}},
\end{equation}
which still scales linearly with the wavelength $\lambda$. 
Then, we assume that the telescope jitter also follows a Gaussian distribution, and thus the irradiation at the telescope focal plane (FP) is:
  \begin{equation}
\label{eq:jit}
E_{jitter}=exp \Big(-\frac{y_{FP}^2}{2(\sigma_{jitter}F_{tel})^2}\Big).
\end{equation}

Early estimates of the jitter amplitude range from $\sigma_{jitter}\geq300 \mu arcsec$ to a few arcseconds\cite{Carrier25,alice2026}. Therefore, the Pollux input PSF is a convolution of the two Gaussian curves. Then, we calculate the PSF radius corresponding to $1\%$ of the irradiation. This value $R_{99\%}$ for the longest wavelength of each channel defines the pinhole radius. The pinhole will crop the PSF at the level of at least $1\%$ and perform the spatial filtering, but will not define the actual input PSF for our instrument. To quantify the input PSF we compute the full width at the half-maximum (FWHM) for the convolved Gaussian. Nevertheless, we are considering implementing the option of multiple pinhole entrances with different sizes to also enable spectrophotometric observations stable over timescales of hours.        

For this specific work, we also assume that all channels will use $\delta$-doped CMOS sensors. The most promising option is to use the CIS 300 platform\cite{Pratlong23}, so the maximum reachable format will be $9k\times8k\times10\mu m$ pixels. The aim is to use only one sensor in each channel. The MUV and NUV channels use the entire $9k\times8k$ area, while the FUV is limited to $8k\times1k$ since it has fewer orders. At the same time it is strictly necessary to guarantee Nyquist sampling of the PSF. So for each channel the dispersion and magnification should be sufficient to have $FWHM\geq2 pix=20\mu m$. In the FUV with its relatively narrow working band it is possible to reach the safety margin of $\geq2.5 pix$, while for the other channels we moderate it to $\geq2.2 pix$. Since the PSF width scales with wavelength, this condition applies to the shortwave end of each spectrogram, while at the longwave end the PSF will be over-sampled. This leads to an increase of the dispersion and magnification, although the exact value of the over-sampling changes depending on jitter.

With these limitations and the top-level specs given in Table~\ref{tab:top_specs} we repeat the calculations for each echelle spectrograph (i.e. FUV, MUV, NUV). The spectral resolving power is limited by the echelle diffraction\cite{Valyavin14}, so the theoretical limit is 
  \begin{equation}
\label{eq:R_theo}
R_{echelle}=\frac{2D_{COL}*tan(\gamma)}{D_{tel}tan(FWHM/f_{tel})},
\end{equation}
where $D_{COL}$ is the collimated beam diameter, $\gamma$ is the echelle blazing angle, which is identical to the incidence angle for the Littrow mount, $D_{tel}$ and $f_{tel}$ are the diameter of telescope's primary mirror and the focal length of the telescope, respectively. In the examples below we consider an optimistic scenario of $D_{tel}=8 m$ and $f_{tel}/D_{tel}=20$. However, the actual resolving power is defined by the finite width of the input PSF, the linear dispersion of the echelle $\partial \lambda / \partial y'$ and the collimator-to-camera geometric magnification:
\begin{equation}
\label{eq:R_disp}
R_{FWHM}=\frac{\lambda}{FWHM\cdot f_{CAM}/f_{COL} \cdot \partial \lambda / \partial y'},
\end{equation}

This value is much lower than the theoretical limit for all of the cases considered in this work. For the sake of simplicity we set the camera focal length to $f_{CAM}=1500 mm$. This allows us to limit its angular field of view to below $3.5^\circ$, so even a spherical concave grating will have relatively low residual aberrations. Otherwise a more challenging CD design would be required.  
The key design parameters for each case are shown in the summary Table~\ref{tab:FUV vs jitter} below.

\begin{figure} [ht]
   \begin{center}
   \begin{tabular}{c} 
   \includegraphics[width=0.9\textwidth]{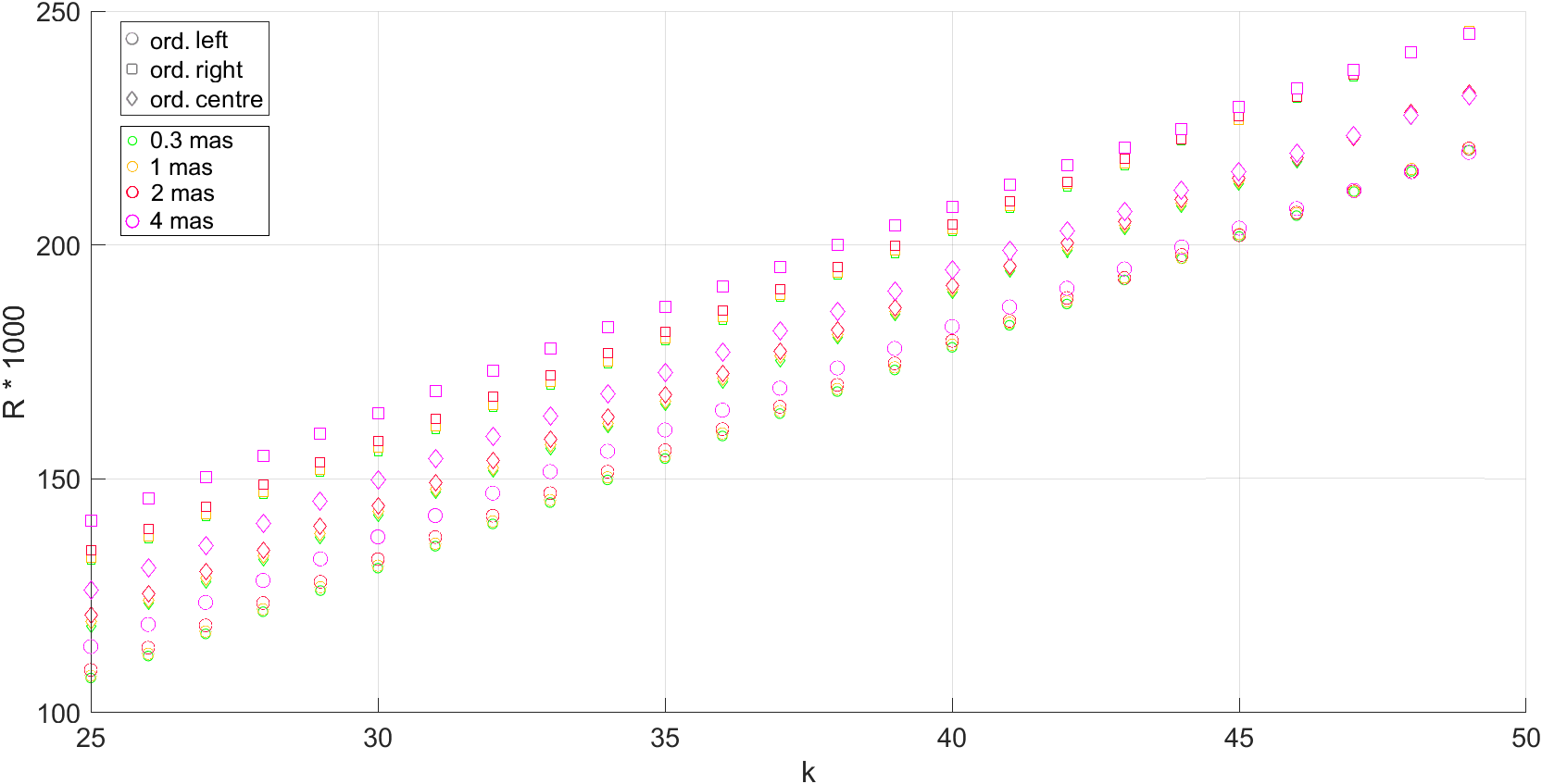}
   \end{tabular}
   \end{center}
   \caption{\label{fig:Resolving_power_MUV}  Spectral resolving power changing across the working orders for the MUV channel. The colors correspond to the telescope pointing jitter amplitude.} 
\end{figure}

The resolving power variation across the working band for the MUV channel is shown in Fig.~\ref{fig:Resolving_power_MUV}, where $k$ is the diffraction order number.

The plot indicates that the resolving power has nearly a linear variance across the orders and a clear linear change in each order. At the same time, the higher jitter values correspond to a higher resolving power for lower orders (i.e. for longer wavelengths), while at higher orders the values are identical. As one can see from Table~\ref{tab:FUV vs jitter}, this also corresponds to a growing de-magnification.   

\begin{figure} [ht]
   \begin{center}
   \begin{tabular}{c} 
   \includegraphics[width=0.85\textwidth]{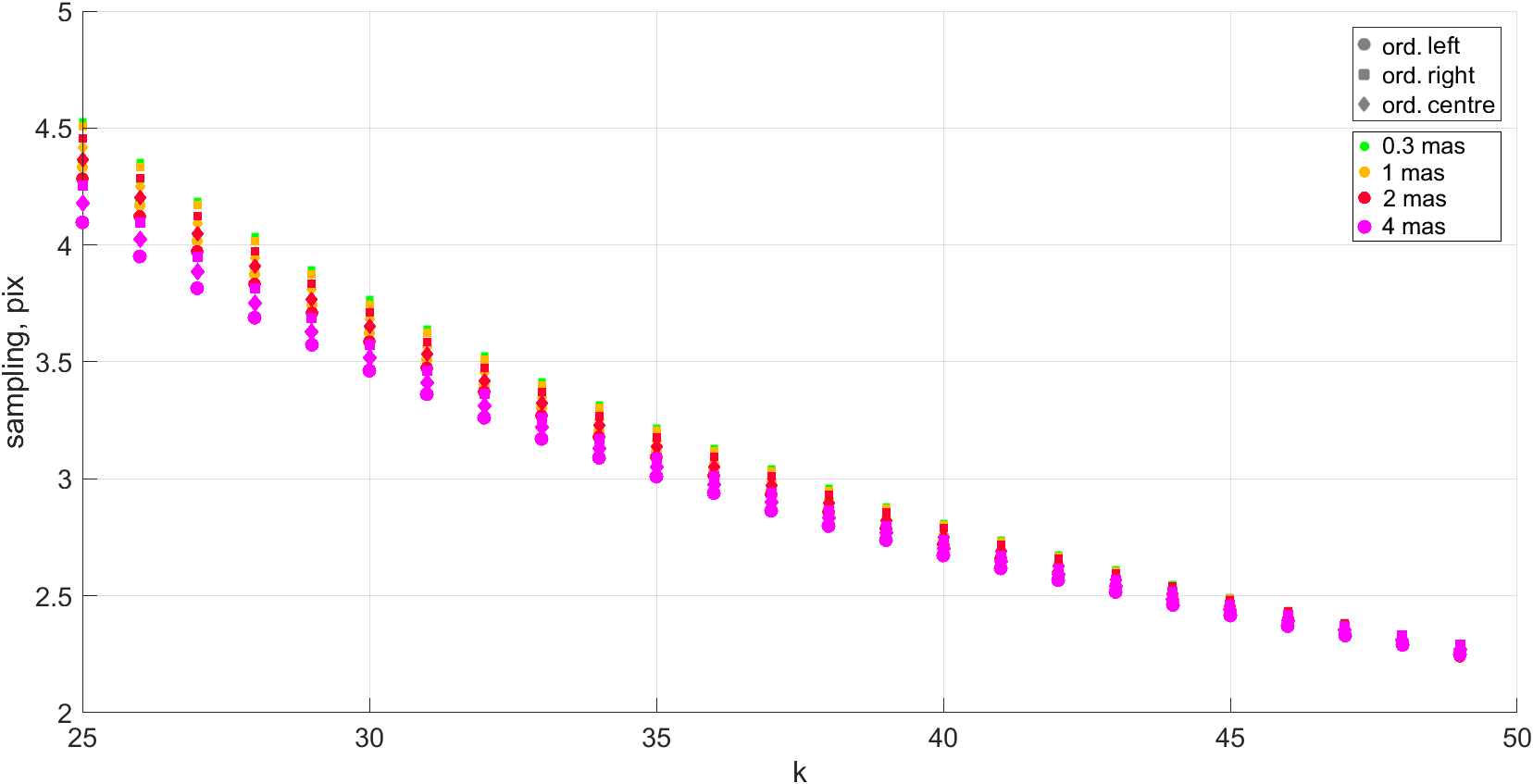}
   \end{tabular}
   \end{center}
   \caption{\label{fig:sampling_MUV}  Spectral resolving power changing across the working orders for the MUV channel.} 
\end{figure}

In a similar way we calculate the sampling for each of these scenarios. The typical plot for the MUV channel is shown in Fig.~\ref{fig:sampling_MUV} and the extreme values are given in Table~\ref{tab:FUV vs jitter}. 

It is clear that for the shortest wavelengths the sampling remains just above the required value. The over-sampling at the highest orders is moderated when a large jitter is assumed, though this effect is relatively small and does not exceed $5.8\%$, so we should not expect any significant moderation of the over-sampling unless the PSF becomes completely dominated by the jitter. 

\begin{figure} [ht]
   \begin{center}
   \begin{tabular}{c} 
   \includegraphics[width=0.85\textwidth]{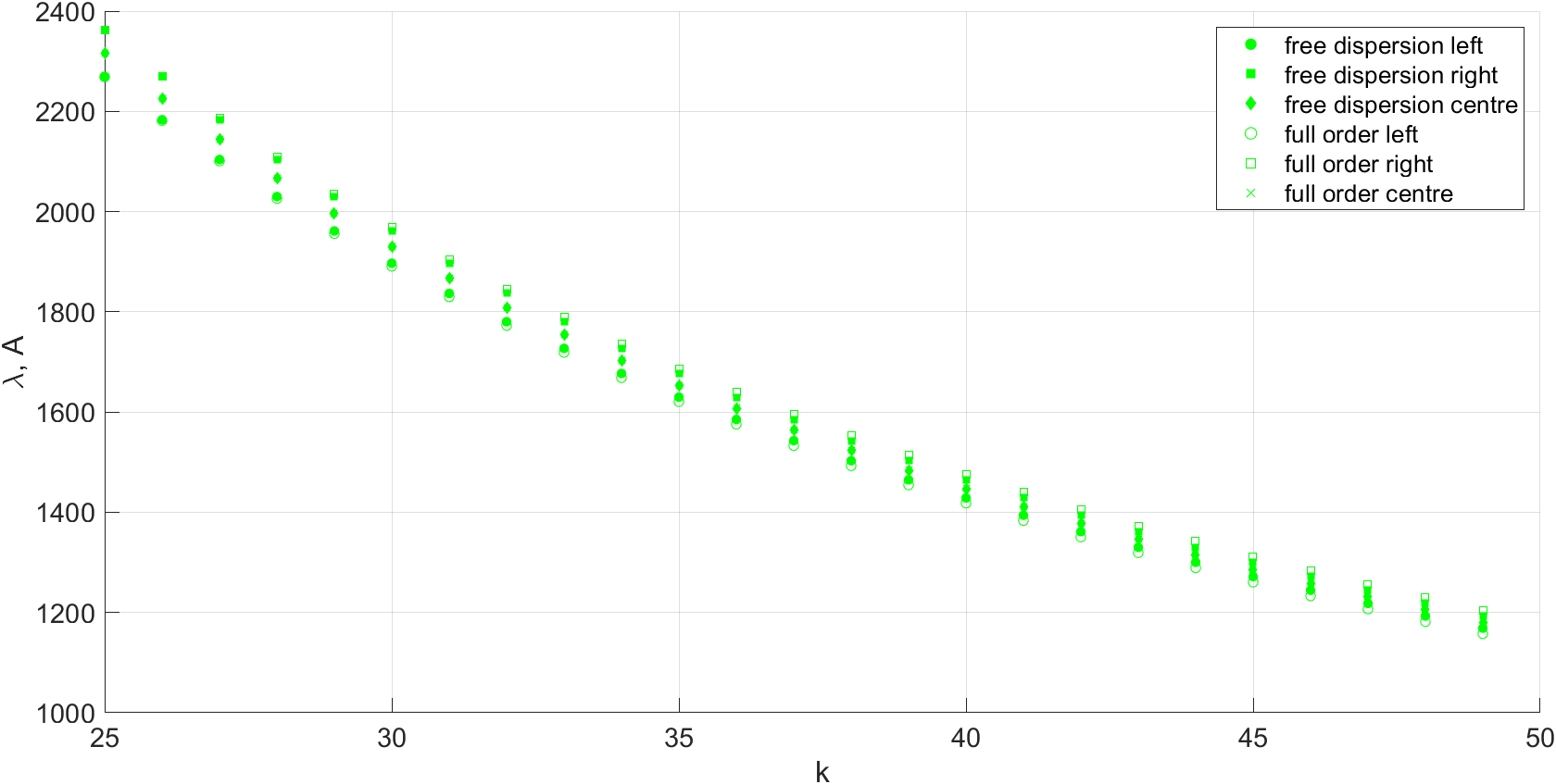}
   \end{tabular}
   \end{center}
  \caption{\label{fig:orders_MUV}  Variance of the full order length and the free dispersion range for the MUV channel.} 
   \end{figure}

Finally, Fig.~\ref{fig:orders_MUV} shows the order length and the free dispersion range varying across the orders. The dependence is clearly non-linear, but the order length is large enough to minimize stitching and avoid corresponding errors.   

In general, if the image quality and stability of the telescope will be as high as expected, the input PSF size will scale linearly with wavelength. This will lead to unavoidable over-sampling at longer wavelengths and apply some additional pressure on the gratings design and the optics packaging. This effect can be slightly moderated by a larger jitter of the telescope, but the qualitative dependence will remain for a large range of input values; in the given example the jitter amplitude varies by more than an order of magnitude.     

\begin{table}[ht]
\caption{\label{tab:FUV vs jitter} Key parameters of the echelle spectrographs and their relation with jitter.}
\begin{center}
\begin{tabular}{lcccc}
\hline
 $\omega_{jitter}, mas$ & 4 & 2 & 1 & 0.3 \\
\hline
\multicolumn{5}{c}{\textbf{FUV}}\\
\hline
$L_{sp}, mm$ & 81.5 & 81.5 & 81.5 & 81.5 \\
\hline
$R_{FWHM}$ & 123 360 & 120 760 & 119 720 & 119 560\\
\hline
$Min. samp., pix$ & 2.39 & 2.37 & 2.37 & 2.36 \\
\hline
$Max. samp., pix$ & 2.96 & 3.05 & 3.07 & 3.08 \\
\hline
$R_{echelle}$ & 20357000 & 19797000 & 19661000 & 19584000 \\
\hline
$N_{echelle}, mm^{-1}$ & 339.0 & 339.0 & 339.0 & 339.0\\
\hline
$\gamma, ^{\circ}$ & 26.7 & 26.7 & 26.7 & 26.7 \\
\hline
$F_{col}, mm$ & 2364 & 2166 & 2112 & 2094 \\
\hline
$k_{min}$ & 27 & 27 & 27 & 27 \\
\hline
$k_{max}$ & 34 & 34 & 34 & 34  \\
\hline
$N_{CD}, mm^{-1}$ & 274.1 & 274.1 & 274.1 & 274.1 \\
\hline
 \multicolumn{5}{c}{\textbf{MUV}}\\
 \hline
 $L_{sp}, mm$& 91.8 & 91.8 & 91.8 & 91.8  \\
\hline
$R_{FWHM}$& 114 140 & 109 160 & 107 920 & 107 420\\
\hline
$Min. samp., pix$ & 2.25 & 2.24 & 2.24 & 2.25 \\
\hline
$Max. samp., pix$ & 4.25 & 4.46 & 4.51 & 4.53 \\
\hline
$R_{echelle}$ & 7470500 & 7130900 & 7048500 & 7015300 \\
\hline
$N_{echelle}, mm^{-1}$ & 174.1 & 174.1 & 174.1 & 174.1\\
\hline
$\gamma, ^{\circ}$ & 14.8 & 14.8 & 14.8 & 14.8 \\
\hline
$F_{col}, mm$ & 2980 & 2798 & 2752 & 2733 \\
\hline
$k_{min}$ & 25 & 25 & 25 & 25 \\
\hline
$k_{max}$& 49 & 49 & 49 & 49  \\
\hline
$N_{CD}, mm^{-1}$ & 480.9 & 480.9 & 480.9 & 480.9 \\
\hline
 \multicolumn{5}{c}{\textbf{NUV}}\\
\hline
$L_{sp}, mm$& 91.6 & 91.6 & 91.6 & 91.6  \\
\hline
$R_{FWHM}$& 112 840 & 111 650 & 111 260 & 111 070\\
\hline
$Min. samp., pix$ & 2.25 & 2.25 & 2.25 & 2.25 \\
\hline
$Max. samp., pix$ & 4.20 & 4.25 & 4.27 & 4.28 \\
\hline
$R_{echelle}$ & 8137800 & 8044600 & 8016400 & 7997000 \\
\hline
$N_{echelle}, mm^{-1}$& 94.8 & 94.8 & 94.8 & 94.8\\
\hline
$\gamma, ^{\circ}$ & 15.9 & 15.9 & 15.9 & 15.9 \\
\hline
$F_{col}, mm$ & 5511 & 5420 & 5393 & 5380  \\
\hline
$k_{min}$ & 25 & 25 & 25 & 25 \\
\hline
$k_{max}$& 46 & 46 & 46 & 46  \\
\hline
$N_{CD}, mm^{-1}$ & 277.9 & 277.9 & 277.9 & 277.9 \\
\hline
\end{tabular}
\end{center}
\end{table}

At the same time, the intention to use only a single sensor in each channel leads to a solution with large de-magnification. When the CD focal length is fixed, the collimator focal length becomes long specially in the NUV channel. This can be resolved either by using double-reflection collimators (i.e. folding mirrors of Cassegrain-alike solutions) or by employing an aberration-corrected CD that works with a wider angular field of view. In the latter case, one can use toroidal or aspheric surfaces, or vary the grating grooves spacing. Another notable point common to all of these design versions is the small blazing angle of the echelles. These values are smaller than typical blazing angles for commercial gratings, so their technological feasibility requires further investigation.

We remark that at this stage of the project there is still some margin for changing the waveband limits of the different channels. For instance, the MUV and NUV ranges could be extended to a full octave each, so they would become $120-240~nm$ and $240-480~nm$, respectively. Obviously, for the MUV the difference is minor, while for the NUV the band extension makes the optical design more challenging. If we apply the $4~mas$ jitter assumption, it would require adding an extra four diffraction orders to cover the whole band, while the echelle grating spatial frequency would decrease to $85.3~mm^{-1}$ and the blazing angle would become even shallower $\gamma=14.5^\circ$. The collimator focal length also increases to $f_{COL}=5602~mm$ and the CD frequency becomes $N_{CD}=231.7 mm^{-1}$. Therefore, extending the working band to the full octave would make the design more challenging and it is advised to keep the current values as long as it is suitable for the science goals.

\section{RETRACTABLE POLARIMETERS}
\label{sec:POLAR}  

Another decision that can affect the instrument architecture and operation modes is the polarimetes retractability. The FUV channel is the one that would benefit the most from implementing a retractable polarimeter because of its low transmission, but the FUV polarimeter is also the most difficult to remove from the beam because of it's size and the beam path in the K-mirror. The goal is to avoid using precise focusing mechanisms as it was done in some other space missions such as EUCLID \cite{Wallner2017}.

\begin{figure} [ht]
   \begin{center}
   \begin{tabular}{c} 
   \includegraphics[width=0.9\textwidth]{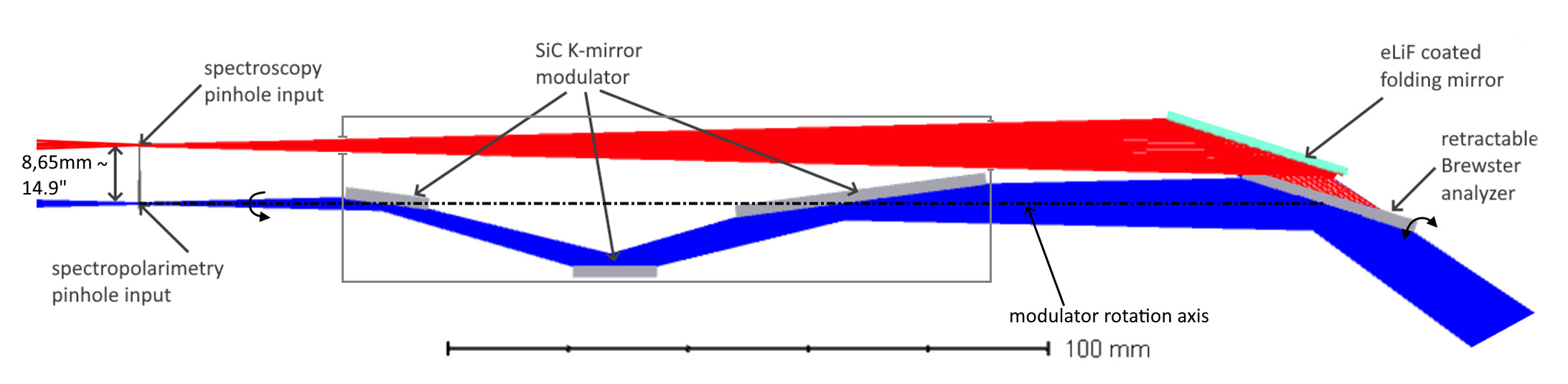}
   \end{tabular}
   \end{center}
 \caption{\label{fig:FUV retracting}  Schematic of the FUV polarimeter by-pass. } 

   \end{figure}

However, there is a geometric solution based on by-passing the polarimeter instead of retracting it in which the FUV modulator has a "parking position" and the input beam can pass through the unit, to be then re-directed to the spectrograph by a fold mirror. In this case the optical path remains the same and there is no need to re-focus the spectrograph (see Fig.~\ref{fig:FUV retracting}).    

\begin{figure} [ht]
   \begin{center}
   \begin{tabular}{c} 
   \includegraphics[width=0.65\textwidth]{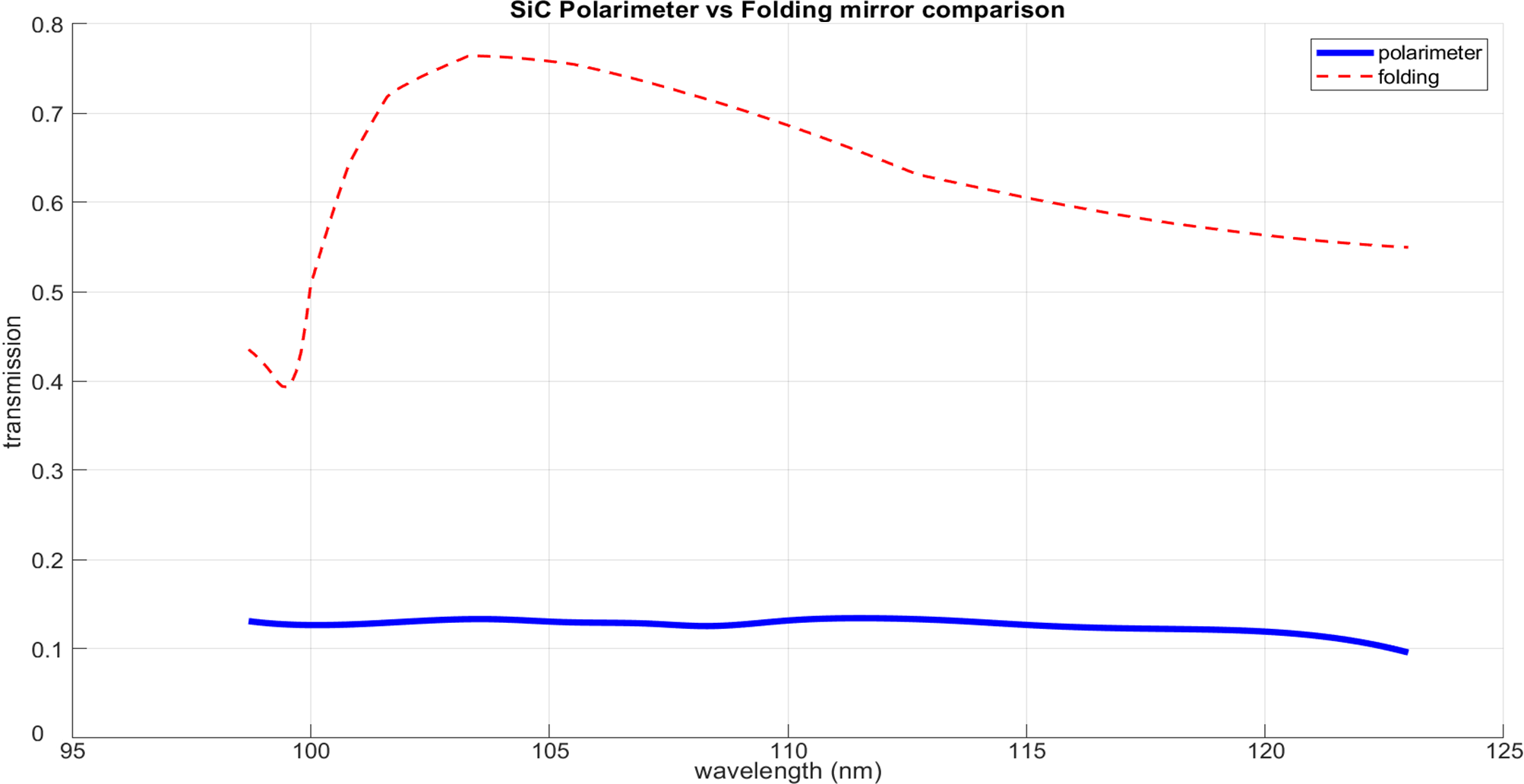}
   \end{tabular}
   \end{center}
    \caption{\label{fig:fuv transmission} Change of the FUV polarimeter unit transmission resulting from by-passing the SiC mirrors comprising the K-mirror FUV polarimeter.} 
   
   \end{figure}

This solution leads to a substantial gain in transmission as Fig.~\ref{fig:fuv transmission} shows. 

Nevertheless, this option requires a few changes in the polarimetric unit design:
\begin{itemize}
    \item The distance between the first modulator mirror and the pinhole increases to $70 mm$ and that between the K-mirror components increases to $35 mm$ each. This means that the technologically challenging SiC mirrors should be slightly bigger: for instance, the last mirror of the K-mirror polarimeter grows in size by $\approx 1.34\%$ to $63.5\times8~mm^2$. 
    \item To enable a FUV spectroscopy mode the entire telescope should be re-pointed by $\approx14.9"$. As we discussed in Sec.~\ref{sec:BASIC}, this should be possible, but adds some complexity to the instrument operations.
    \item There is the need to implement a precise mechanism to remove the Brewster analyzer from the beam. However, such mechanism will have just two positions for which it is easier to reach the required high precision, e.g. by use of a mechanical stop. 
\end{itemize}
   
The change in transmission is evident as four SiC mirrors in the optical train are replaced by one mirror with an optimized LiF-based coating\cite{DeMarcos18, Moore14}. As the plot shows, the maximum gain in transmission can reach $\approx5.9^\times$. A further advantage of this design is that it enables the implementation of a slit-spectroscopy mode also in the FUV channel, though this would require adding a further mechanism. 

    \begin{figure} [hbt]
   \begin{center}
   \begin{tabular}{c} 
   \includegraphics[width=0.57\textwidth]{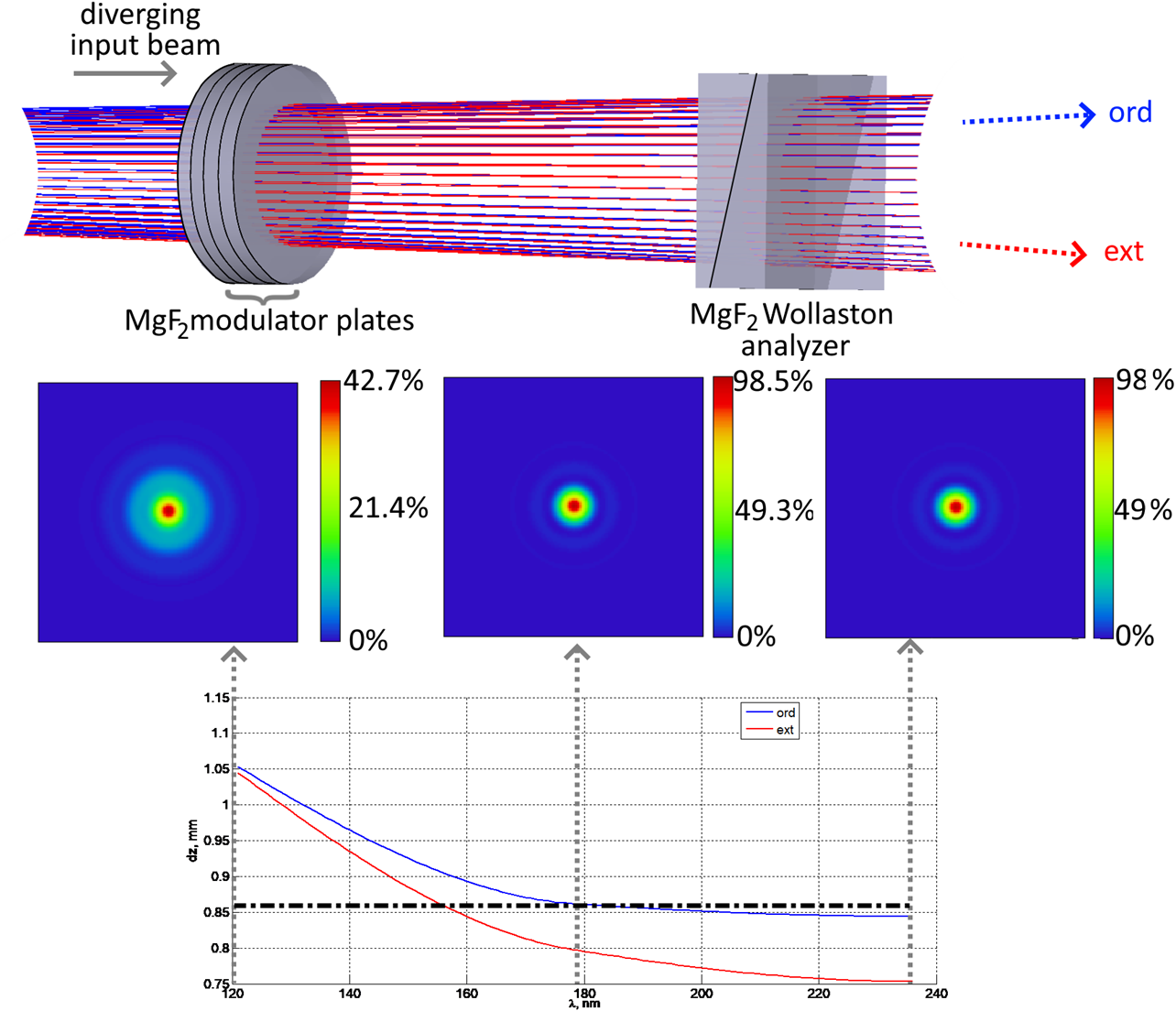}
   \end{tabular}
   \end{center}
    \caption{\label{fig:mgf2 defocus} Birefringent polarimeter unit for the MUV channel: top -- optical design, middle -- chromatic change of the PSF and the corresponding Strehl ratios, bottom --  longitudinal chromatism for the ordinary and extraordinary beams.} 
    \end{figure} 
    
For the NUV and MUV channels it is possible to use birefringent materials, and  in particular $MgF_2$ crystals. The design of these polarimeters is based on thin plates of $1.2 mm$ only in total thickness and a Wollaston prism to separate the components with a thickness of $1.5 mm$. However, the difference in optical path following the insertion of the polarimeter is non-negligible and requires re-focusing.  In addition, the $MgF_2$ elements are chromatic, so the defocus varies across the band. Fig.~\ref{fig:mgf2 defocus} shows the polarimeter design, the longitudinal defocus and chromatic change of the PSF.   

   \begin{figure} [ht]
   \begin{center}
   \begin{tabular}{c} 
   \includegraphics[width=0.57\textwidth]{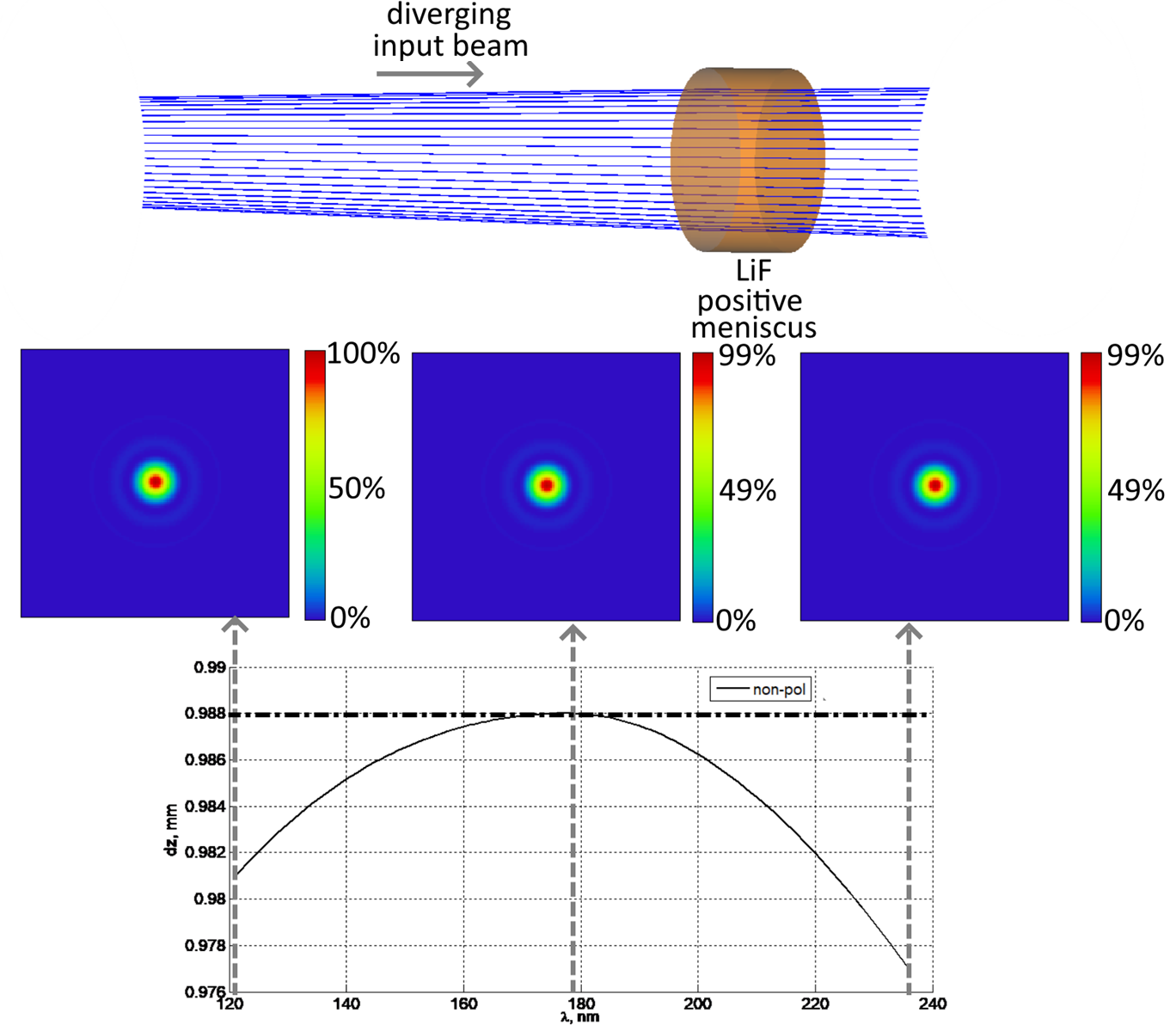}
   \end{tabular}
   \end{center}
  \caption{\label{fig:lif refocus} Polarimeter retraction compensator for the MUV channel: top -- optical design, middle -- chromatic change of the PSF and the corresponding Strehl ratios, bottom --  longitudinal chromatism for the ordinary and extraordinary beams.} 
   \end{figure} 
   
To avoid the need to re-focus the detector, it is possible to introduce a compensator lens that re-images the pinhole to the same position. For the MUV channel the choice of transparent materials is limited. $LiF$ crystal represent a good option in terms of transmission and chromatic properties. In this case, the compensator represents a slow meniscus that is placed instead of the Wollaston prism, while the modulator is removed. This solution does not add any complexity to the retracting mechanism and can even help with its balancing.  
Fig.~\ref{fig:lif refocus} shows the compensator design and its nearly-achromatic properties.

 \begin{figure} [htb]
   \begin{center}
   \begin{tabular}{c} 
   \includegraphics[width=0.7\textwidth]{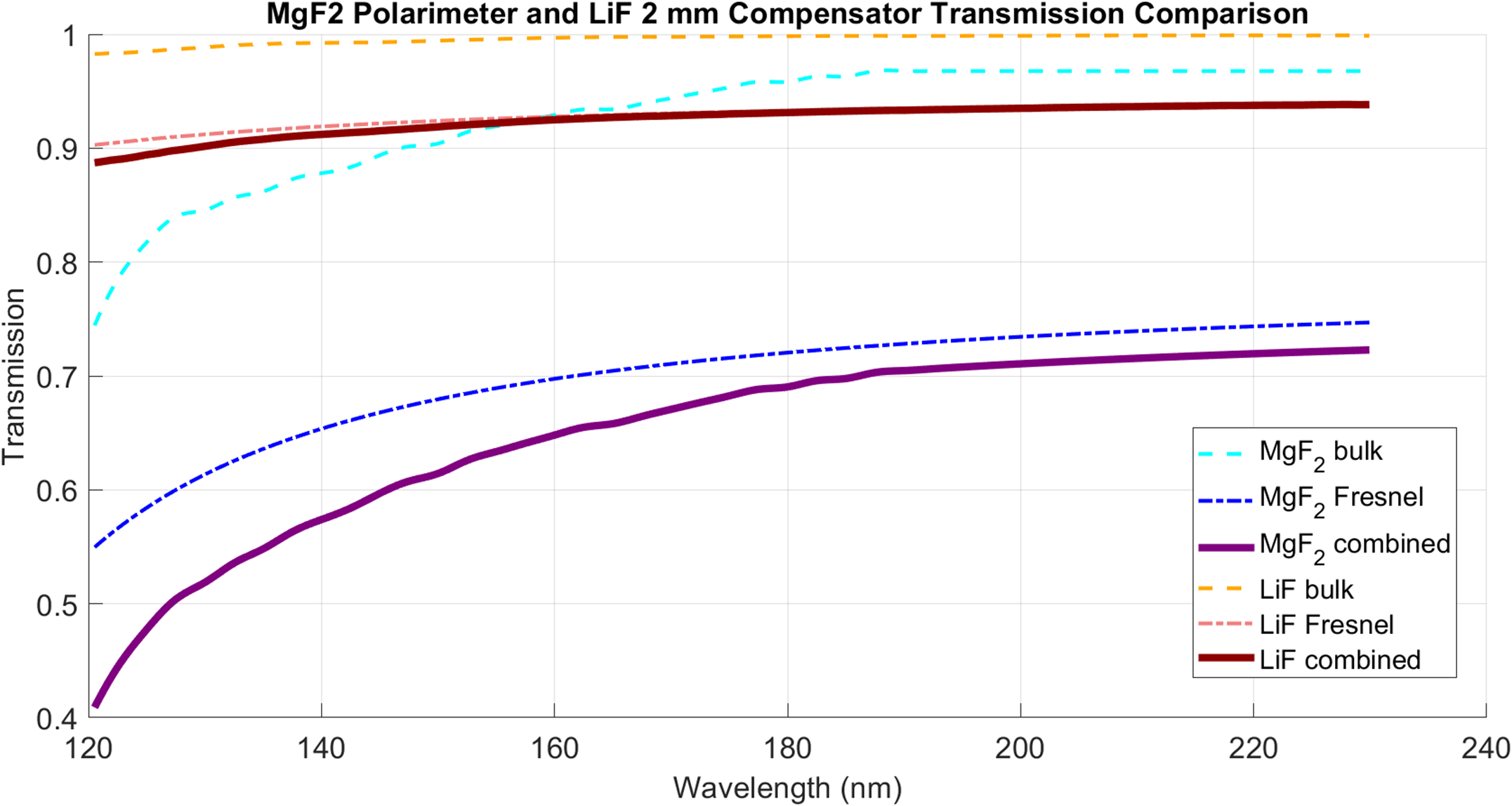}
   \end{tabular}
   \end{center}
\caption{\label{fig:birefring transmission} Comparison of the spectral transmission between the MUV $MgF_2$-based birefringent polarimeter and the $LiF$ compensator lens. } 
   \end{figure} 
   
Therefore, by using this compensator lens we can increase the image quality and avoid a re-focusing mechanism. At the same time the transmission gain is notable, especially at the shortwave end. If we consider uncoated elements of $MgF_2$ and $LiF$, the difference in transmission for the highest diffraction order can reach $\approx2.2^\times$ (Fig.~\ref{fig:birefring transmission}).
   
It may be difficult to maintain the polarization properties after applying an anti-reflective (AR) coating on the polarization elements, so in this case the transmission is defined by the Fresnel losses. However, even if the UV AR coating is possible and the transmission is defined only by the bulk absorption, the $LiF$ \cite{Li1976} lens has a clear benefit in comparison to the $MgF_2$ \cite{Dodge84} elements. The NUV compensator can be designed in a similar way, although the gain in transmission will be lower. Also, the NUV compensator can be made of conventional optical materials such as UV-grade fused silica or $CaF_2$, and thus less challenging.

   \section{CONCLUSIONS}
\label{sec:CONC}  

This work led us to the following conclusions:
\begin{itemize}
    \item It is likely that the telescope PSF width will scale with wavelength in a wide range of residual pointing jitter amplitudes. Therefore, given the very wide Pollux wavelength coverage the PSF size variation will likely be large, and thus an over-sampling of up to $2^\times$ must be introduced in the optical design.  
    \item It is possible to fit the high-resolution spectra within each UV channel on single CIS300-series CMOS sensors. However, this will have implications for the collimator or camera designs, as well as for the echelle diffraction geometry. 
    \item There is a promising solution to by-pass the FUV polarimeter, thus enabling a high-efficiency spectroscopy mode without the need to re-focus the detector. 
    \item Similarly, the use of a compensator lens enables one to implement the retraction of the MUV and NUV birefringent polarimeters without the need to adopt a complicated re-focusing system. Such a lens can be nearly-achromatic and for the shortwave part of the MUV band it has up to $2^\times$ higher transmission.    
\end{itemize}

These solutions will affect the top-level architecture of Pollux. However, the optical and mechanical designs will likely need to be adapted to match the telescope interfaces that are yet to be defined.

\acknowledgments 
 
We acknowledge the continued support of HWO-Pollux studies provided by CNES through APR grants since 2022.

\bibliography{report} 

@inproceedings{Pratlong23,
author = {Jerome Pratlong},
title = {{Innovations in visible imaging technology to support future space missions}},
volume = {12777},
booktitle = {International Conference on Space Optics — ICSO 2022},
editor = {Kyriaki Minoglou and Nikos Karafolas and Bruno Cugny},
organization = {International Society for Optics and Photonics},
publisher = {SPIE},
pages = {1277712},
year = {2023},
doi = {10.1117/12.2689043},
URL = {https://doi.org/10.1117/12.2689043}
}

@misc{Feinberg26,
      title={Habitable Worlds Observatory's Concept and Technology Maturation: Initial Feasibility and Trade Space Exploration}, 
      author={Lee D. Feinberg and Breann N. Sitarski and Michael W. McElwain and Giada Arney and Caleb Baker and Matthew R. Bolcar and Marie Levine and Alice Liu and Bertrand Mennesson and Aki Roberge and J. Scott Smith and Feng Zhao and John Ziemer},
      year={2026},
      eprint={2601.11803},
      archivePrefix={arXiv},
      primaryClass={astro-ph.IM},
      url={https://arxiv.org/abs/2601.11803}, 
}

@inproceedings{Girardot24,
author = {Adrien Girardot and Coralie Neiner and Jean-Michel Reess},
title = {{Design of a FUV polarimeter for Pollux aboard HWO}},
volume = {13093},
booktitle = {Space Telescopes and Instrumentation 2024: Ultraviolet to Gamma Ray},
editor = {Jan-Willem A. den Herder and Shouleh Nikzad and Kazuhiro Nakazawa},
organization = {International Society for Optics and Photonics},
publisher = {SPIE},
pages = {130933V},
year = {2024},
doi = {10.1117/12.3017994},
URL = {https://doi.org/10.1117/12.3017994}
}

@article{DeMarcos18,
author = {Luis V. Rodr\'{i}guez De Marcos and Juan I. Larruquert and Jos\'{e} A. M\'{e}ndez and Nuria Guti\'{e}rrez-Luna and Luc\'{i}a Espinosa-Y\'{a}\~{n}ez and Carlos Honrado-Ben\'{i}tez and Jos\'{e} Chavero-Roy\'{a}n and Bel\'{e}n Perea-Abarca},
journal = {Opt. Express},
number = {7},
pages = {9363--9372},
publisher = {Optica Publishing Group},
title = {Optimization of MgF2-deposition temperature for far UV Al mirrors},
volume = {26},
month = {Apr},
year = {2018},
url = {https://opg.optica.org/oe/abstract.cfm?URI=oe-26-7-9363},
doi = {10.1364/OE.26.009363}
}

@inproceedings{ Moore14,
author = {Christopher Samuel Moore and John Hennessy and April D. Jewell and Shouleh Nikzad and Kevin France},
title = {{Recent developments and results of new ultraviolet reflective mirror coatings}},
volume = {9144},
booktitle = {Space Telescopes and Instrumentation 2014: Ultraviolet to Gamma Ray},
editor = {Tadayuki Takahashi and Jan-Willem A. den Herder and Mark Bautz},
organization = {International Society for Optics and Photonics},
publisher = {SPIE},
pages = {91444H},
year = {2014},
doi = {10.1117/12.2057272},
URL = {https://doi.org/10.1117/12.2057272}
}

@article{Dodge84,
author = {Marilyn J. Dodge},
journal = {Appl. Opt.},
number = {12},
pages = {1980--1985},
publisher = {Optica Publishing Group},
title = {Refractive properties of magnesium fluoride},
volume = {23},
month = {Jun},
year = {1984},
url = {https://opg.optica.org/ao/abstract.cfm?URI=ao-23-12-1980},
doi = {10.1364/AO.23.001980},
}

@inproceedings{Neiner25,
author = {Coralie Neiner and Adrien Girardot and Jean-Michel Reess},
title = {{Space UV polarimeters}},
volume = {13699},
booktitle = {International Conference on Space Optics — ICSO 2024},
editor = {Fr{\'e}d{\'e}ric Bernard and Nikos Karafolas and Philippe Kubik and Kyriaki Minoglou},
organization = {International Society for Optics and Photonics},
publisher = {SPIE},
pages = {1369928},
year = {2025},
doi = {10.1117/12.3072789},
URL = {https://doi.org/10.1117/12.3072789}
}

@ARTICLE{Valyavin14,
       author = {{Valyavin}, G.~G. and {Bychkov}, V.~D. and {Yushkin}, M.~V. and {Galazutdinov}, G.~A. and {Drabek}, S.~V. and {Shergin}, V.~S. and {Sarkisyan}, A.~N. and {Semenko}, E.~A. and {Burlakova}, T.~E. and {Kravchenko}, V.~M. and {Kudryavtsev}, D.~O. and {Pritychenko}, A.~M. and {Kryukov}, P.~G. and {Semjonov}, S.~L. and {Musaev}, F.~A. and {Fabrika}, S.~N.},
        title = "{High-resolution fiber-fed echelle spectrograph for the 6-m telescope. I. Optical scheme, arrangement, and control system}",
      journal = {Astrophysical Bulletin},
         year = 2014,
        month = apr,
       volume = {69},
       number = {2},
        pages = {224-239},
          doi = {10.1134/S1990341314020102},
       adsurl = {https://ui.adsabs.harvard.edu/abs/2014AstBu..69..224V}
}

@misc{Neiner26,
      title={The Pollux European instrument concept for HWO: a high-resolution spectrograph and spectropolarimeter from the far-UV to the near-IR}, 
      author={Coralie Neiner and Jean-Claude Bouret and Luca Fossati and David le Mignant and Eduard Muslimov and Ana Ines Gomez de Castro and Frédéric Marin},
      year={2026},
      eprint={2602.09828},
      archivePrefix={arXiv},
      primaryClass={astro-ph.IM},
      url={https://arxiv.org/abs/2602.09828}, 
}

@inproceedings{LeMignant26,
author = {David LeMignant and Coralie Neiner and Jean-Claude Bouret and Luca Fossati},
title = {{Pollux: high-resolution precision spectroscopy and polarimetry for the Habitable Worlds Observatory}},
volume = {14146},
booktitle = {Space Telescopes and Instrumentation 2026: Ultraviolet to Gamma Ray},
organization = {International Society for Optics and Photonics},
publisher = {SPIE},
year = {2026},
}

@INPROCEEDINGS{Wallner2017,
       author = {{Wallner}, Oswald and {Ergenzinger}, Klaus and {Tuttle}, Sean and {Vaillon}, L. and {Johann}, Ulrich},
        title = "{EUCLID mission design}",
    booktitle = {Society of Photo-Optical Instrumentation Engineers (SPIE) Conference Series},
         year = 2017,
       series = {Society of Photo-Optical Instrumentation Engineers (SPIE) Conference Series},
       volume = {10565},
        month = nov,
          eid = {105650K},
        pages = {105650K},
          doi = {10.1117/12.2309226},
       adsurl = {https://ui.adsabs.harvard.edu/abs/2017SPIE10565E..0KW}
}

@misc{alice2026,
      title={Early Architecture Concepts for the Habitable Worlds Observatory -- System Design, Modeling, and Analysis}, 
      author={Alice and Liu and Marie Levine and Charley Noecker and Jon Lawrence and Joshua Abel and Michael Akkerman and Eric Aanstaat and Ruslan Belikov and Pin Chen and Kenneth Dziak and Jordan Effron and Lee Feinberg and Alan Gostin and James Govern and Cameron Haag and Joseph Howard and Brian Kern and Gary Kuan and Milan Mandic and Carson McDonald and Connor Mulrenin and Bijan Nemati and Jon Papa and Fang Shi and Samuel Sirlin and Breann Sitarski and Cory Smiley and J. Scott Smith and Philip Stahl and Christopher Stark and Gregory Walsh and John Ziemer},
      year={2026},
      eprint={2602.11046},
      archivePrefix={arXiv},
      primaryClass={astro-ph.IM},
      url={https://arxiv.org/abs/2602.11046}, 
}

@inproceedings{Muslimov24,
author = {Eduard Muslimov and Coralie Neiner and Jean-Claude Bouret},
title = {{Optical design options for Pollux: UV spectropolarimeter project for the Habitable Worlds Observatory}},
volume = {13093},
booktitle = {Space Telescopes and Instrumentation 2024: Ultraviolet to Gamma Ray},
editor = {Jan-Willem A. den Herder and Shouleh Nikzad and Kazuhiro Nakazawa},
organization = {International Society for Optics and Photonics},
publisher = {SPIE},
pages = {130933D},
year = {2024},
doi = {10.1117/12.3020175},
URL = {https://doi.org/10.1117/12.3020175}
}

@inproceedings{Carrier25,
author = {Alain Carrier and Kiarash Tajdaran and Michael S. Jacoby and Alison Nordt and Liz Osborne},
title = {{Ultrastability evaluation of pathfinder architecture for Habitable Worlds Observatory}},
volume = {13623},
booktitle = {UV/Optical/IR Space Telescopes and Instruments: Innovative Technologies and Concepts XII},
editor = {Jonathan W. Arenberg and H. Philip Stahl},
organization = {International Society for Optics and Photonics},
publisher = {SPIE},
pages = {1362304},
year = {2025},
doi = {10.1117/12.3067621},
URL = {https://doi.org/10.1117/12.3067621}
}

@ARTICLE{Li1976,
       author = {{Li}, H.~H.},
        title = "{Refractive index of alkali halides and its wavelength and temperature derivatives}",
      journal = {Journal of Physical and Chemical Reference Data},
         year = 1976,
        month = apr,
       volume = {5},
       number = {2},
        pages = {329-528},
          doi = {10.1063/1.555536},
       adsurl = {https://ui.adsabs.harvard.edu/abs/1976JPCRD...5..329L}
}
\bibliographystyle{spiebib} 

\end{document}